\documentclass[aps,prl,twocolumn,superscriptaddress,showpacs]{revtex4-1}

\usepackage{graphicx}
\usepackage{amsmath,amssymb}
\usepackage{bm}
\usepackage{color}

\makeatletter
\def\btt#1{\texttt{\@backslashchar#1}}%
\DeclareRobustCommand\bblash{\btt{\@backslashchar}}%
\makeatother

\begin{document}


\title{Fate of a soliton matter upon symmetry-breaking ferroelectric order}

\author{K. Sunami}
\email{e-mail: sunami@mdf2.t.u-tokyo.ac.jp}
\affiliation{Department of Applied Physics, University of Tokyo, Bunkyo-ku, Tokyo 113-8656, Japan}

\author{R. Takehara}
\affiliation{Department of Applied Physics, University of Tokyo, Bunkyo-ku, Tokyo 113-8656, Japan}

\author{A. Katougi}
\affiliation{Department of Applied Physics, University of Tokyo, Bunkyo-ku, Tokyo 113-8656, Japan}

\author{K. Miyagawa}
\affiliation{Department of Applied Physics, University of Tokyo, Bunkyo-ku, Tokyo 113-8656, Japan}

\author{S. Horiuchi}
\affiliation{Research Institute for Advanced Electronics and Photonics (RIAEP), National Institute of Advanced Industrial Science and Technology (AIST), Tsukuba, Ibaraki, 305-8565, Japan}

\author{R. Kato}
\affiliation{Condensed Molecular Materials Laboratory, RIKEN, Wako, Saitama, 351-0198, Japan}

\author{T. Miyamoto}
\affiliation{Department of Advanced Materials Science, University of Tokyo, Kashiwa, Chiba, 277-8561, Japan}

\author{H. Okamoto}
\affiliation{Department of Advanced Materials Science, University of Tokyo, Kashiwa, Chiba, 277-8561, Japan}
\affiliation{AIST-UTokyo Advanced Operando-Measurement Technology Open Innovation Laboratory (OPERANDO-OIL), National Institute of Advanced Industrial Science and Technology (AIST), Chiba 277-8568, Japan}

\author{K. Kanoda}
\email{e-mail: kanoda@ap.t.u-tokyo.ac.jp}
\affiliation{Department of Applied Physics, University of Tokyo, Bunkyo-ku, Tokyo 113-8656, Japan}

\date{\today}

\begin{abstract}
In a one-dimensional (1D) system with degenerate ground states, their domain boundaries, dubbed solitons, emerge as topological excitations often carrying unconventional charges and spins; however, the soliton excitations are only vital in the non-ordered 1D regime. Then a question arises; how do the solitons conform to a 3D ordered state? Here, using a quasi-1D organic ferroelectric, TTF-CA, with degenerate polar dimers, we pursue the fate of a spin-soliton charge-soliton composite matter in a 1D polar-dimer liquid upon its transition to a 3D ferroelectric order by resistivity, NMR and NQR measurements. We demonstrate that the soliton matter undergoes neutral spin-spin soliton pairing and spin-charge soliton pairing to form polarons, coping with the 3D order. The former contributes to the magnetism through triplet excitations whereas the latter carries electrical current. Our results reveal the whole picture of a soliton matter that condenses into the 3D ordered state.
\end{abstract}


\maketitle

Coupling between charge, spin and lattice in solids gives rise to emergent low-energy excitations, which appear as solitons of topological nature in one-dimensional (1D) systems with degenerate ground states \cite{Heeger_1988, Brazovskii_2006}. The quasi-1D organic donor-acceptor complex, tetrathiafulvalene-$p$-chloranil (abbreviated as TTF-CA) [Fig. 1(a)] offers an exclusive ground for the physics of charge solitons and spin solitons \cite{Nagaosa_1986, Soos_2007, Fukuyama_2016, Tsuchiizu_2016}. The TTF and CA molecules are nominally neutral at ambient temperature and pressure, however, pressurizing or cooling induces a neutral-to-ionic (NI) transition \cite{Torrance_1981, Torrance_1981_1, Dressel_2017, Masino_2017, Avino_2017, Cointe_2017, Takehara_2018} with an abrupt charge transfer from TTF to CA; the schematic phase diagram is shown Fig. 1(b). At low temperatures [the green-colored region in Fig. 1(b)], the ionic state is accompanied by a lattice dimerization (donor-acceptor pairing) due to the spin-Peierls instability \cite{Girlando_1983, Tokura_1985}, yielding a non-magnetic symmetry-breaking electronic ferroelectric ($I_\mathrm{ferro}$) phase taking either of two degenerate dimerization patterns [Fig. 1(a)] \cite{Cointe_1995, Kobayashi_2012}. The ferroelectric dimer order melts into a polar dimer liquid upon entering a paraelectric ionic ($I_\mathrm{para}$) phase at high temperatures [the orange-colored region in Fig. 1(b)] \cite{Takehara_2018, Cailleau_1997}.

\begin{figure}
\includegraphics[width=8.5cm,clip]{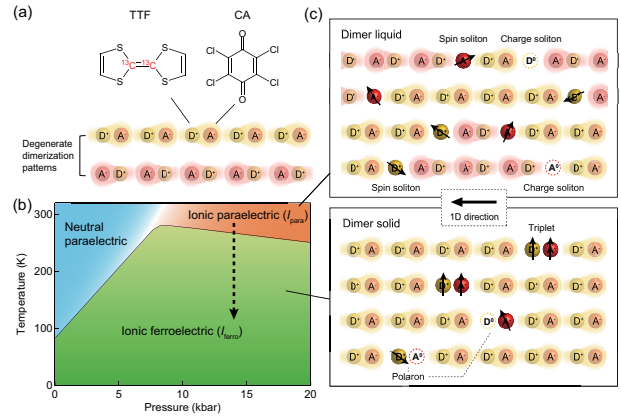}
\caption{Phase diagram and spin and charge excitations in TTF-CA. (a) Molecular structures of TTF and CA. The central double-bonded carbon atoms in TTF molecule are enriched by $^{13}$C isotopes for $^{13}$C-NMR measurements. Two degenerate dimerization patterns with opposite polarizations are illustrated at the bottom. D and A represent the donor and acceptor molecules, TTF and CA, respectively. (b) Pressure-temperature phase diagram of TTF-CA. The broken arrow indicates the trace of measurements in the present study. (c) Illustrations of spin and charge excitations in the ionic paraelectric and ferroelectric phases of TTF-CA, respectively.
}
\label{Fig1} 
\end{figure}

\begin{figure*}
\includegraphics[width=13cm, clip]{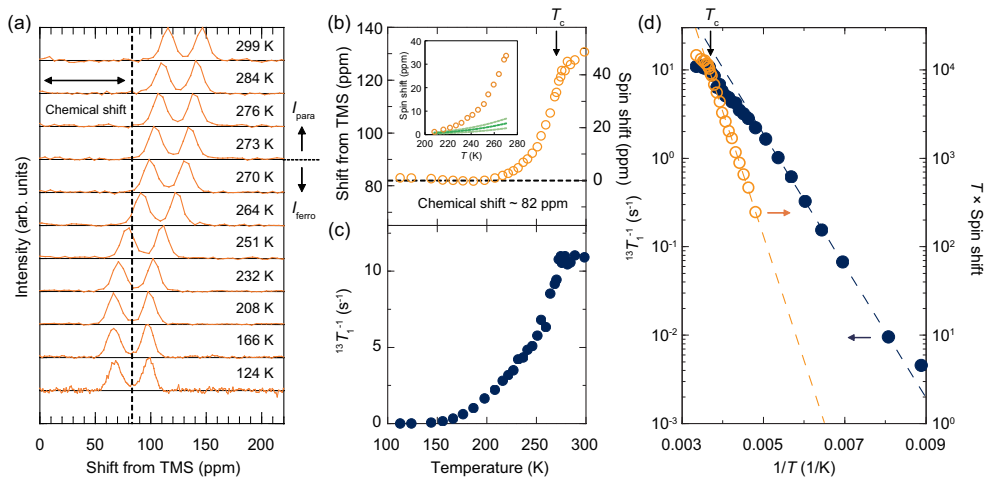}
\caption{$^{13}$C-NMR spectra, shift and relaxation rate. (a) Temperature dependence of $^{13}$C-NMR spectra at 14 kbar. The doublet structure arises from the $^{13}$C-$^{13}$C nuclear dipolar coupling and the NMR shift is given by the midpoint of the doublet. The shift origin corresponds to the resonance frequency of TMS (tetramethylsilane). (b) Plot of the midpoint of the doublet as a function of temperature. Left and right axes represent the total shift and the spin shift [= total shift $-$ chemical shift (82 ppm)], respectively. Inset: Zoom-up of the behavior near $T_\mathrm{c}$. Estimates of polaron contribution to spin shift are indicated by three green lines (upper and lower limits, and their median) (see text and Supplemental Material \cite{SM} for the details of the estimation).  (c) Temperature dependence of $^{13}$C-NMR spin-lattice relaxation rate $^{13}$$T_1^{-1}$ at 14 kbar. (d) Comparison between $^{13}$$T_1^{-1}$ (blue closed circles; left axis) and $T \times$(spin shift) (orange open circles; right axis) plotted against inverse temperature. The blue broken line is a fit of the single exponential function to $^{13}$$T_1^{-1}$ in 120 $< T <$ 200 K. The orange broken line is a fit of the single exponential function to $T \times$(spin shift) in 200 K $< T <$ $T_\mathrm{c}$.
}
\label{Fig2}
\end{figure*}

In the $I_\mathrm{para}$ phase, the space inversion symmetry is globally preserved but locally broken \cite{Okamoto_1989, Masino_2007} such that $S$ = 1/2 spin solitons and spinless charge solitons \cite{Nagaosa_1986, Soos_2007, Fukuyama_2016, Tsuchiizu_2016} are thermally excited to interrupt the global order and generate oppositely polarized dimer domains [Fig. 1(c)]. At 14 kbar and at ambient temperature, for example, the spin soliton density is one per 10-25 donor-acceptor pairs, as revealed by a recent nuclear magnetic resonance (NMR) study \cite{Sunami_2018}, whereas the charge-soliton density is one per $\sim$100 donor-acceptor pairs according to an analysis of a transport experiment (Supplemental Material \cite{SM}). A soliton matter comprised of spin solitons (majority) and charge solitons (minority) resides in the $I_\mathrm{para}$ phase. An issue of profound interest but yet addressed in soliton physics is what happens in the soliton matter upon entering a 3D ordered phase, which does not allow free soliton excitations to violate the 3D order \cite{Karpov_2016}. The soliton matter should not be able to preserve the pristine state. The present study gives a solution to this fundamental issue by investigating TTF-CA under temperature variation across the $I_\mathrm{para}$ and $I_\mathrm{ferro}$ phases through electrical conductivity, $^{13}$C-NMR and $^{35}$Cl-NQR (nuclear quadrupole resonance) measurements probing charge, spin and lattice, respectively.

For $^{13}$C-NMR measurements, we synthesized $^{13}$C-enriched TTF molecules, in which the central double-bonded carbon atoms are labelled by $^{13}$C isotopes with a 99\% concentration in the method described in Ref. \cite{Sunami_2018}. Both $^{13}$C-enriched and non-enriched single crystals of TTF-CA were prepared by a co-sublimation method. Hydrostatic pressure was applied to the sample using a nonmagnetic BeCu clamp-type cell (for $^{13}$C-NMR and $^{35}$Cl-NQR) and a BeCu/NiCrAl dual-structured one (for electrical conductivity measurement) with Daphne 7373 oil as the pressure medium. Electrical conductivity was measured with electrical currents applied along the $a$ axis (1D direction) of a single crystal by the four-terminal method. $^{13}$C-NMR and $^{35}$Cl-NQR measurements were conducted under an external magnetic field of 8 tesla directed to the $a$ axis and under zero field, respectively. The signals of nuclear magnetization were obtained using the solid-echo pulse sequence for $^{13}$C-NMR and the spin-echo pulse sequence for $^{35}$Cl-NQR. The nuclear spin-lattice relaxation rate is determined by fitting the single exponential function to the relaxation curve of nuclear magnetization.

\begin{figure*}
\includegraphics[width=13cm]{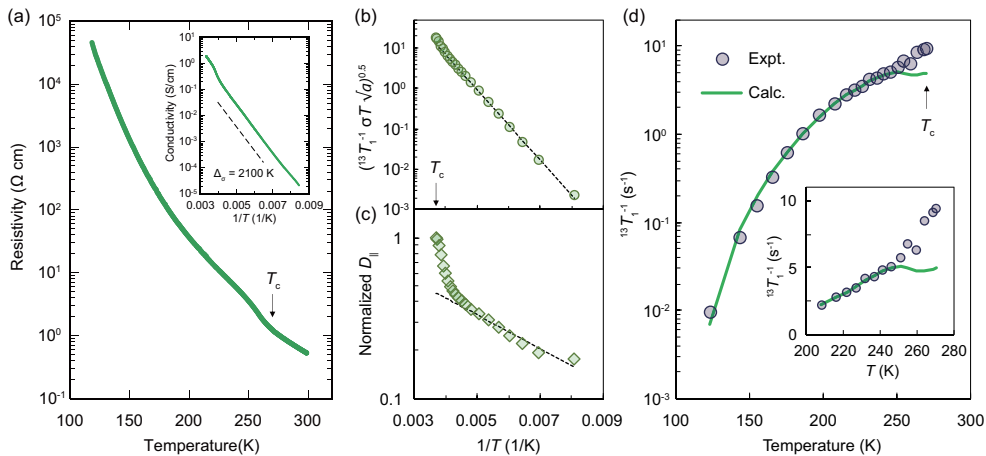}
\caption{Electrical resistivity. (a) Temperature dependence of electrical resistivity at 14 kbar. Inset: Activation plot of electrical conductivity, $\sigma$. The dotted line represents the Arrhenius form with the activation energy of 2100 K. (b) Plot of $(^{13}T_1^{-1}\sigma T\sqrt{a})^{0.5}$, which is proportional to the density of polarons $n$ well below $T_\mathrm{c}$ (see text). The dotted line represents the Arrhenius form with the activation energy of 2010 K estimated using the data below 250 K. (c) Temperature dependence of diffusion constant evaluated through Eq. (3) using $\sigma$ and $n$ (see text); the value at $T_\mathrm{c}$ is normalized to unity. The dotted line represents the Arrhenius form with the activation energy of 240 K estimated using the data below 220 K. (d) Comparison between the observed $^{13}$C-NMR spin-lattice relaxation rate $^{13}T_1^{-1}$ and the contribution of polarons to $^{13}T_1^{-1}$, $(^{13}T_1^{-1})_\mathrm{p}$, calculated using $\sigma$ (see text). $(^{13}T_1^{-1})_\mathrm{p}$ is normalized to $^{13}T_1^{-1}$ at 190 K. Inset: Behavior near $T_\mathrm{c}$ in linear scales. The excess in $^{13}T_1^{-1}$ from the $(^{13}T_1^{-1})_\mathrm{p}$ curve near $T_\mathrm{c}$ is likely the contribution of the triplet excitations of the bound soliton pairs.
}
\label{Fig3} 
\end{figure*}

First, we conducted the $^{13}$C-NMR measurements with temperature varied across the ferroelectric transition under a pressure, 14 kbar. At every temperature studied, $^{13}$C-NMR spectra have two peaks [Fig. 2(a)] that arise from nuclear dipolar interactions between the central $^{13}$C sites. As temperature is lowered, the spectral shift decreases with a clear kink at a transition temperature, $T_\mathrm{c}$ $\sim$ 270 K, in accordance with the previous NQR result \cite{Takehara_2018}, and saturates to the value of 82 ppm [Fig. 2(b)]. The spectral shift is the sum of the spin shift, $S$, proportional to the spin susceptibility, and the chemical shift caused by the orbital motion of electrons. The $I_\mathrm{ferro}$ phase is non-magnetic due to the spin-singlet formation \cite{Sunami_2018, Mitani_1984} so that we take the saturated value, 82 ppm, as the chemical shift. The plot of $S$ (= the observed shift $-$ 82 ppm) multiplied by temperature, $T$, vs. 1/$T$ [Fig. 2(d)] is well characterized by $TS$ $\propto$ exp($-\Delta_s$/$k_\mathrm{B}T$) with a spin excitation gap, $\Delta_s$, of 3240 K for 200 K $< T <$ $T_\mathrm{c}$, where $k_\mathrm{B}$ is the Boltzmann constant. The conventional spin-Peierls systems are known to hold the relationships between $T_\mathrm{c}$ and the singlet-triplet gap $\Delta$, $\Delta$/$k_\mathrm{B}$$T_\mathrm{c}$ $\sim$ 1.76 (the BCS relationship) or 2.47 (obtained by a bosonization method \cite{Orignac_2004}). The present value, $\Delta_s$/$k_\mathrm{B}$$T_\mathrm{c}$ is $\sim$12, is too large for the conventional spin-Peierls picture for the 1D Heisenberg spins. Indeed, the $I_\mathrm{para}$ phase carries mobile spin and charge solitons, qualitatively different from the conventional paramagnetic phases \cite{Sunami_2018, Takehara_2019}. In the $I_\mathrm{ferro}$ phase below $T_\mathrm{c}$, the soliton excitations should be in pairs to keep the 3D ferroelectric order, and thus the $\Delta_s$ of 3240 K characterizes the excitations of the triplet-type neutral spin soliton-antisoliton pairs. The order parameter of the dimer order develops only in a narrow temperature width of 10-20 K below $T_\mathrm{c}$ (Supplemental Material \cite{SM}); thus the pair strongly bound in a DA dimer becomes loosely separated only near $T_\mathrm{c}$ and unbound above $T_\mathrm{c}$.

Unconventional feature of spin excitations is also captured by the $^{13}$C nuclear spin-lattice relaxation rate, $^{13}$$T_1^{-1}$, which is nearly independent of temperature above $T_\mathrm{c}$ but decreases below $T_\mathrm{c}$ with a kink at $T_\mathrm{c}$ as in $S$ [Fig. 2(c)]. The activation plot of $^{13}$$T_1^{-1}$ exhibits an exponential decrease characterized by a gap value of $\Delta_{T_1^{-1}}$ = 1720 K defined by $T_1^{-1}$ $\propto$ exp($-\Delta_{T_1^{-1}}$/$k_\mathrm{B}T$) in 120 $< T <$ 200 K, whereas the variation of $^{13}$$T_1^{-1}$ is gradual in 200 K $< T <$ $T_\mathrm{c}$, where the slope of activation plot of $^{13}$$T_1^{-1}$ is much less than that of $S$ with $\Delta_s$ = 3240 K [Fig. 2(d)]. In the conventional singlet-triplet excitations, the spin excitation gaps in $S$ and $T_1^{-1}$ should not significantly differ \cite{Itoh_1997}; for the spin-Peierls case, the activation energy of $T_1^{-1}$ is theoretically predicted to be twice that of $S$ in $T$ $\ll$ $\Delta$/$k_\mathrm{B}$ due to the indirect three magnon process \cite{Ehrenfreund_1977}. In TTF-CA, however, the observed moderate decrease of $^{13}$$T_1^{-1}$ is totally unexplainable by this process, suggesting the presence of another form of spin excitations below $T_\mathrm{c}$. If a spin soliton and a charge soliton are bound to form a ``polaron'' with an elementary charge and a spin 1/2 \cite{Tsuchiizu_2016}, it can be excited and carry charges and spins without violating the ferroelectric dimer order in the $I_\mathrm{ferro}$ phase [Fig. 1(c)]. As we discuss later, the polaron excitation is a major contributor to $^{13}$$T_1^{-1}$ except near $T_\mathrm{c}$ whereas triplet excitation of bound spin soliton pairs is so to $S$ near $T_\mathrm{c}$.

The electrical resistivity at 14 kbar is insulating below room temperature with a slight kink at $T_\mathrm{c}$ [Fig. 3(a)]; the activation energy of conductivity $\sigma$ in the $I_\mathrm{ferro}$ phase is $\Delta_\sigma$ $\sim$ 2100 K, which appears not much changed in the $I_\mathrm{para}$ phase [inset of Fig. 3(a)]. Quasiparticle excitations would give a transport activation energy of larger than the half of the optical charge-transfer gap of $\sim$0.7 eV ($\sim$8100 K) in the ionic phase \cite{Torrance_1981, Okamoto_2004}. The substantially lower value of $\Delta_\sigma$ $\sim$ 2100 K (or 2$\Delta_\sigma$ $\sim$ 4200 K for a particle-antiparticle creation) suggests low-energy charge excitations distinct from quasiparticles. Note that charge soliton excitations, which violate a 3D order, are prohibited in the $I_\mathrm{ferro}$ phase but allowed when attaching themselves to spin solitons to form polarons, which is the most likely case that explains the present observation.

Consequently, there would be two types of spin excitations in the $I_\mathrm{ferro}$ phase; the triplet-type bound spin soliton pairs and the polaronic bound pairs of spin and charge solitons. In the $I_\mathrm{para}$ phase at room temperature and 14 kbar, the spin-soliton density is one per 10-25 donor-acceptor pairs \cite{Sunami_2018} whereas the charge-soliton density is one per $\sim$100 donor-acceptor pairs as mentioned above (Supplemental Material \cite{SM}); namely, the majority is the spin solitons. Thus, the magnetism just below $T_\mathrm{c}$ should be dominated by the bound spin solitons, whose triplet excitations with the large gap ($\Delta_s$ of 3240 K) cause the steep decrease in spin shift. Well below $T_\mathrm{c}$, where the triplet excitations almost vanish, the polarons with the lower excitation gaps ($\Delta_{T_1^{-1}}$ of 1720 K in $^{13}$$T_1^{-1}$ and $\Delta_\sigma$ of 2100 K in $\sigma$) would be main contributors of magnetism and conductivity. The following analysis based on $^{13}$$T_1^{-1}$ and $\sigma$ gives further insight into the polaron formation.

Given that polarons move diffusively, its contribution to $^{13}T_1^{-1}$, $(^{13}T_1^{-1})_\mathrm{p}$, is expressed as (Supplemental Material \cite{SM}), 
\begin{equation}
(^{13}T_1^{-1})_\mathrm{p} \propto n/\sqrt{D_{\parallel}D_{\perp}},
\end{equation}
where $n$ is the density of polarons, and $D_{\parallel}$ ($D_{\perp}$) is the diffusion constant of polarons along the direction parallel (perpendicular) to the 1D chains. Eq. (1) is rewritten in terms of the temperature-dependent anisotropy parameter, $a(T) = D_{\perp}/D_{\parallel}$, as
\begin{equation}
(^{13}T_1^{-1})_\mathrm{p} \propto n/\sqrt{a}D_{\parallel}.
\end{equation}
On the other hand, $\sigma$ is expressed through the Einstein relation as 
\begin{equation}
\sigma = ne^2D_{\parallel}/k_\mathrm{B}T,
\end{equation}
where $e$ is the elementary charge. Eqs. (2) and (3) yield $n \propto [(^{13}T_1^{-1})_\mathrm{p}\sigma T\sqrt{a}]^{0.5}$, which is evaluated using the experimental values of $^{13}T_1^{-1}$ and $\sigma$ and the anisotropy of conductivity measuring the $a$ values (Supplemental Material \cite{SM}). As shown in Fig. 3(b), $n$ obeys $n \propto \mathrm{exp}(-\Delta_n/k_\mathrm{B}T)$ with $\Delta_n$ $\sim$ 2010 K; a slight deviation of $n$ from the activation line near $T_\mathrm{c}$ maybe an artifact arising from the contribution of the bound spin solitons to $^{13}T_1^{-1}$. Applying the deduced activation form of $n$ to the conductivity formula, Eq. (3), we obtain the temperature variation of $D_{\parallel}$ as shown in Fig. 3(c). At low temperatures, $D_{\parallel}$ shows an activation behavior indicating the thermal hopping over the energy barrier of 240 K. Remarkably, this energy scale is near the Peierls-coupled optical phonon frequencies $\sim$ 120-180 K in the $I_\mathrm{ferro}$ phase \cite{Masino_2006}, suggesting that the polarons diffuse assisted by the Peierls phonon modes. The $D_{\parallel}$ goes up from the activation line upon approaching $T_\mathrm{c}$ most likely because the Peierls modes critically enhanced near $T_\mathrm{c}$ promotes the diffusion of the polarons. Then, substituting the obtained $D_{\parallel}$ and the activation form of $n$ to Eq. (2), we evaluate the polaron contribution to the relaxation rate, $(^{13}T_1^{-1})_\mathrm{p}$, which nearly coincides with the experimental $^{13}T_1^{-1}$ values up to 250 K [Fig. 3(d)]. This suggests that the polaron motions are responsible for $^{13}T_1^{-1}$ in 120 $< T <$ 250 K; the deviation of the experimental $^{13}T_1^{-1}$ values from $(^{13}T_1^{-1})_\mathrm{p}$ in 250 K $< T <$ $T_\mathrm{c}$ is likely the contribution of the triplet excitations of the bound soliton pairs to $^{13}T_1^{-1}$. We also evaluate the polaron contribution to spin shift with reference to $n$ and the densities of spin and charge solitons in the $I_\mathrm{para}$ phase [see the inset of Fig. 2(b)]; it shows that the triplets dominate the spin shift down to $\sim$230 K, below which the two contributions are comparable. This estimation is in agreement with the interpretation that the triplet contribution is captured in $^{13}T_1^{-1}$ just below $T_\mathrm{c}$.

\begin{figure}
\includegraphics[width=8.3cm]{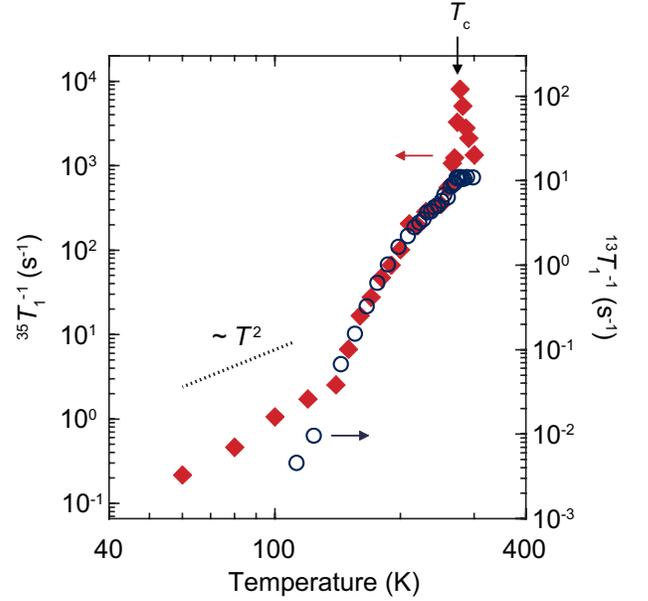}
\caption{$^{35}$Cl-NQR relaxation rate. Comparison between $^{35}$Cl-NQR spin-lattice relaxation rate $^{35}T_1^{-1}$ (red closed diamonds; left axis) and $^{13}T_1^{-1}$ (blue open circles; right axis) under 14 kbar. The dotted line indicates the $T^2$ law expected from the conventional phonons.
}
\label{Fig4}
\end{figure}

In order to get further evidence for the polaron excitations, we measured $^{35}$Cl-NQR $T_1^{-1}$, $^{35}T_1^{-1}$ at 14 kbar, which probes the lattice fluctuations though nuclear quadrupolar interaction. $^{35}T_1^{-1}$ exhibits a divergent peak at $T_\mathrm{c}$ and decreases with temperature (Fig. 4). The peak in $^{35}T_1^{-1}$, in sharp contrast to its absence in $^{13}T_1^{-1}$, is attributable to the critical lattice fluctuations associated with the 3D ferroelectric dimerization transition and clearly indicates that $^{35}T_1^{-1}$ probes quadrupole relaxation instead of magnetic relaxation through hyperfine interaction. This is consistent with the absolute value of $^{35}T_1^{-1}$ that is too large to interpret in terms of the hyperfine interaction (Supplemental Material \cite{SM}). At low temperatures below 140 K, $^{35}T_1^{-1}$ is roughly proportional to $T^2$, suggesting the conventional phonon contribution (two-phonon Raman process) to the nuclear quadrupole relaxation \cite{Abragam_1961}. Above 140 K, however, another relaxation contribution appears and, remarkably, $^{35}T_1^{-1}$ and $^{13}T_1^{-1}$ show the common temperature evolution for 140 $< T <$ 250 K ($< T_\mathrm{c}$), indicating that the quadrupolar and magnetic relaxations have a common origin. This strongly supports that the polarons carrying spins travel accompanied by local lattice distortion, which should cause quadrupole and magnetic relaxations with the same temperature profiles.

In the present work, we tackled the problem of how spin solitons and charge solitons vitally excited in a 1D polar dimer liquid conform into a 3D ferroelectric dimer order in a neutral-ionic transition system, TTF-CA. The NMR, NQR, and conductivity measurements all coherently point to a binding transition of the soliton matter to two-component composite pairings comprised of neutral spin soliton pairs and polaronic spin-soliton charge-soliton pairs. The spin soliton pairs contribute to magnetism through triplet excitations, which rapidly decrease upon cooling, whereas the polarons dominate the low-temperature magnetism and conductivity, and diffusively travel with a hopping activation energy close to the Peierls-coupled optical phonon energies, suggestive of the Peierls phonon-assisted hopping. Solitons are mobile topological defects of keen interest and has been intensively explored particularly in regard to its individual properties. The present work has revealed how the solitons are organized when the system in the 1D regime enters into the 3D symmetry breaking ordered regime, offering a new perspective to soliton physics.

\acknowledgments
We thank H. Fukuyama for fruitful discussions. This work was supported by the JSPS Grant-in-Aids for Scientific Research (grant nos. JP16H06346, JP17K05532, JP18H05225 and JP19H01846), by CREST (grant no. JPMJCR1661), Japan Science and Technology Agency and by the Murata Science Foundation.



\end{document}